# Coalescent embedding in the hyperbolic space unsupervisedly discloses the hidden geometry of the brain

Alberto Cacciola[1,†], Alessandro Muscoloni[2,3,†], Vaibhav Narula[2,3], Alessandro Calamuneri[6], Salvatore Nigro[7], Emeran A. Mayer[9,10,11,12], Jennifer S. Labus[9,10,11], Giuseppe Anastasi[6], Aldo Quattrone[7,8], Angelo Quartarone[1,6], Demetrio Milardi[1,6] and Carlo Vittorio Cannistraci[2,3,4,5,*]

[1]IRCCS Centro Neurolesi "Bonino Pulejo", Messina, Italy.

[2]Biomedical Cybernetics Group, Biotechnology Center (BIOTEC), [3]Center for Molecular and Cellular Bioengineering (CMCB), [4]Center for Systems Biology Dresden, [5]Department of Physics, Technische Universität Dresden, Dresden, Germany.

[6]Department of Biomedical, Dental Sciences and Morphological and Functional Images, University of Messina, Messina, Italy.

[7]Institute of Bioimaging and Molecular Physiology, National Research Council, Catanzaro, 88100, Italy.

[8]Institute of Neurology, Department of Medical and Surgical Sciences, University "Magna Graecia", Catanzaro, 88100, Italy.

[9]G. Oppenheimer Center for Neurobiology of Stress and Resilience, UCLA, Los Angeles, CA, United States

[10]Department of Medicine, UCLA, Los Angeles, CA, United States

[11]UCLA Vatche and Tamar Manoukian Division of Digestive Diseases, UCLA, Los Angeles, CA, United States

[12]UCLA Brain Research Institute, Los Angeles, CA, United States

*Correspondence should be addressed to:
Carlo Vittorio Cannistraci (kalokagathos.agon@gmail.com)

†The first two authors should be regarded as joint First Authors

**The human brain displays a complex network topology, whose structural organization is widely studied using diffusion tensor imaging. The original geometry from which emerges the network topology is known, as well as the localization of the network nodes in respect to the brain morphology and anatomy. One of the most challenging problems of current network science is to infer the latent geometry from the mere topology of a complex network. The human brain structural connectome represents the perfect benchmark to test algorithms aimed to solve this problem.**
*Coalescent embedding* **was recently designed to map a complex network in the hyperbolic space, inferring the node angular coordinates. Here we show that this methodology is able to** *unsupervisedly* **reconstruct the latent geometry of the brain with an incredible accuracy and that the intrinsic geometry of the brain networks strongly relates to the lobes organization known in neuroanatomy. Furthermore, coalescent embedding allowed the detection of geometrical pathological changes in the connectomes of Parkinson's Disease patients. The present study represents the first evidence of brain networks'** *angular coalescence* **in the hyperbolic space, opening a completely new perspective, possibly towards the realization of latent geometry network markers for evaluation of brain disorders and pathologies.**

Studying the brain as a network of interconnected nodes and the recent developments of network theory, contributed to unveil the key structural principles underlying the topology of the healthy human brain [1,2]. One of the peculiar rules on which brain topology relies is the tendency of the network nodes to cluster into modules with high efficiency and short path length, thus reflecting an intrinsic small-world architecture, functionally segregated (local clustering) and integrated (global efficiency) [3–5]. Indeed, structural magnetic resonance (MR) studies based on diffusion tensor imaging (DTI, an imaging technique that allows to perform tractography and connectome reconstruction) have demonstrated that the human brain shows a modularity structure, consisting of around five to six modules, corresponding to known functional subsystems [3,6]. In addition to the existence of a structural core [3], the brain seems to exhibit a rich-club organization, with highly connected and central nodes (hubs) having a strong tendency to be mutually interconnected, thus constituting a focal point for whole-brain communication [7].

Such topological patterns of connectivity are often intricately related with the physical distances between elements in brain networks. On one hand, brain regions that are spatially close have a relatively high probability of being interconnected, on the other hand, longer white matter projections are more expensive in terms of their material and energy costs, thus making connections between spatially far brain structures less likely [8].

Interestingly, the topology of many real networks seems to be characterized by a latent hyperbolic geometry and the hyperbolic space is a promising universal space of representation for real networks, preserving many of their fundamental topological properties [9]. However, mapping a given real network to its hyperbolic space remains still an open challenge. The Popularity Similarity Optimization (PSO) model proposed that the trade-off between node popularity and similarity contributes to establish new connections and the hyperbolic space offers a congruous geometrical representation for this mechanism of self-organization [9]. In particular, according to the PSO model, the radial coordinates and the angular distances of the nodes in the hyperbolic disk respectively represent the node popularity and similarity.

To the best of our knowledge, the few attempts that have been made to reveal the brain connectivity's intrinsic geometry in the Euclidean space failed [10], and the embedding of brain connectomes is still an unexplored field. In this regard, by using game theory and three-dimensional Euclidean embedding, Gulyás et al. [11] showed that the brain has a highly navigable skeleton, which reflects that the spatial organization of the brain is nearly optimal for information transfer. Ye et al. (2015), using dimensionality reduction techniques, one linear (multidimensional scaling) and one nonlinear (Isomap), introduced a new mathematical framework that, as anticipated above, failed to represent the three-dimensional intrinsic geometry of the human brain connectome [10], concluding that such "intrinsic geometry only minimally relates to neuroanatomy" [10]. However, the fact that the intrinsic geometry they inferred did not relate to neuroanatomy could be imputed to embedding limitations of the employed algorithms. In fact, the brain networks are physically expensive systems and it is likely that several features of real brain anatomy have been structured to control such wiring costs [8]. It is well known that human brain networks show both anatomically short-distance connections and long-distance pathways between several spatially remote modules and anatomical areas.

Network geometry is a promising field and has the potential to promote a significant improvement in revealing and understanding the hidden geometry from which emerges a complex network structure in healthy and pathological conditions. Nevertheless, as we mentioned above, mapping real networks into the hyperbolic space remains an open problem. Recently, Thomas et al.[12] and Muscoloni et al. (article under revision)[13] proposed *coalescent embedding*, a class of topological-based unsupervised nonlinear dimension reduction machine learning able to perform efficient mapping of complex networks in the 2D hyperbolic disk, the

3D hyperbolic sphere, and potentially also in higher-dimensions. Applying this new class of algorithms, we investigated whether we were able to demonstrate unsupervisedly - exploiting only the mere topology of the brain networks – that it is possible to disclose in the hyperbolic space the fundamental properties of the hidden brain geometry, which in our case is inferred from the MR-DTI structural connectivity. We considered three different brain weighted networks, each network was obtained from a different dataset (Table 1) as the mean or median of all the weighted networks that characterized the healthy individuals included in the dataset. The network weights of the first and second datasets present a value that relates to the number of streamlines (NOS) between brain regions, while in the third dataset the streamlines distance (SD) is reported. In general, we selected datasets that differ for typology of weights and the technical details are given in the Methods and Suppl. Information.

Coalescent embedding in the 2D hyperbolic disk perfectly segregates the structural networks into two distinct sections, corresponding to the left and right hemispheres (Fig. 1) of the brain. The segregation has value 1 on a range of [0,1] estimated using two different circular-separation-scores (see methods for technical details), which are measures of accuracy that evaluate the level of correct separation of the node labels (in this case left and right) on the angular coordinates of the disk. Such pattern of segregation clearly emerged from all the datasets analysed (Table 1), but for simplicity of narration hereafter we will show the figures of the dataset that offered the best results. In this case, we selected *Dataset I* (Nigro et al.[14]) and the median-network is represented in Fig. 1. It seems that the first rule of organization of brain networks that emerges in the geometrical space is their structural segregation in two hemispheres, which is a simple concept yet quite neglected in previous studies on brain connectomics. Furthermore, we observed that also the anterior, central and posterior part of the brain were correctly allocated and segregated in the 2D hyperbolic disk (Fig. 2). Circular-separation-scores - evaluating the matching between the angular organization of the nodes in the 2D embedding space and their real anatomical arrangement (according to node labels anterior-central-posterior) - are shown in Table 1 for the three datasets analysed. Interestingly, for the *Dataset I*, ncMCE offered also in this case a perfect angular node alignment (Fig. 2). Hence, we can conclude that both left-right and frontal-central-back node arrangement - obtained by coalescent embedding in the 2D hyperbolic disk - respected the brain morphology with an impressive level of accuracy, providing the first important result of this study.

At this point, we were ready to investigate a more complicated hypothesis: if the structural brain networks hide a latent but clear anatomy-related geometry, the result of coalescent embedding in the 2D hyperbolic disk should respect also the traditional and well-known brain lobes organization. A previous study failed to prove this important correspondence [10]. Fig. 3 instead surprisingly confirms such hypothesis, and shows that the brain median-network of the *Dataset I* embedded by ncMCE segregates into spatially and anatomically distinct sub-networks corresponding to the brain lobes with almost perfect matching. Table 1 shows the circular separation scores for each of the three datasets, considering both mean and median networks for each dataset, and using all the coalescent embedding algorithms. Interestingly, the level of matching between embedding and anatomy is always high (accuracy>0.7) for at least one of the coalescent embedding algorithms in all the three datasets. In particular, we notice that ncMCE, which is a hierarchical embedding technique, offers top performance in the first dataset, while manifold techniques such as ISO and ncISO works better in the other two datasets. We speculate that these results might be related with the different strategies and link-weight-variables used to build the connectomes of the three different datasets, and we leave this technical topic open for future studies.

As explained in[12,13], among the unsupervised machine learning techniques adopted for coalescent embedding, the manifold-based (ISO, ncISO and LE) were the only ones that could be extended for mapping to the three-dimensional (3D) hyperbolic space. Although the 2D hyperbolic disk already offers an almost perfect reconstruction of the brain anatomy, the addition of the third dimension highlights the close relation between the latent geometry and the real brain lobes anatomy. Astonishingly, Figs. 4-6 and Suppl. Video 1,2,3 show that regardless of the manifold-based coalescent embedding technique adopted, all of them could reconstruct a 3D brain network mapping that resembles that original lobar geometrical arrangement proper of the known brain anatomy.

In addition, in *Dataset III (van den Heuvel),* we found that human structural brain networks exhibited a significant different geometry in two age range-specific groups (22-25 and 31-35 years old respectively). This latent geometry variation clearly emerged using coalescent embedding and considering as markers for the discrimination the average hyperbolic distance (HD) and hyperbolic shortest path (HSP) between the nodes[12,13]. On the contrary, the average of the original weights could not reveal any significant change between the two groups (Table 2). We also investigated the possibility to detect age-related geometrical changes in the brain networks using the 3D hyperbolic sphere. Suppl. Table 1 confirms that all the methods and hyperbolic markers, apart from LE-3D-HSP, allow to discriminate between the two conditions using the three dimensions. However, there is not a significant improvement with respect to the 2D embedding, which appears to be enough to detect age-related between-groups variation in the latent geometry of brain networks contained in this dataset.

If the age modifies the latent geometry of the healthy connectomes, pathological conditions related with brain degeneration should impair the latent geometry of patients' connectomes. Therefore, it was natural to investigate whether hyperbolic markers such as the ones described above, could be useful tools also to detect pathological-related between-groups variations in *de novo* drug naïve Parkinson's Disease (PD) patients in respect to healthy controls. The last interesting finding of our work is that our algorithms allow the detection of brain network geometrical pathological differences in the hyperbolic space. Indeed, we demonstrate that both the HD and HSP markers allow to uncover the altered latent geometry in the structural brain networks of the PD patients. On the contrary, also in this comparison, the mean of the weights in the original topology could not provide significant differences between the controls and PD groups (Table 3). Then, we widened the framework from the 2D to the 3D hyperbolic space. The analysis of the 3D hyperbolic embedding revealed that all the coalescent embedding methods and all the hyperbolic markers could significantly identify the alteration in the intrinsic brain geometry of *de novo* drug naïve PD patients as shown in Suppl. Table 2. However, again the 3D embedding did not offer anything more in comparison to the 2D.

Before to conclude we took also in consideration the option to adopt HyperMap [15] and HyperMap-CN [16] (two alternative methods for mapping in the hyperbolic disk) for the group-comparisons. It was not possible to run them on *Dataset III (van den Heuvel)* since the number of networks was too big. Given the high computational complexity of these methods, we could apply them on *Dataset IV (Cacciola)* only. Suppl. Table 3 shows that just HyperMap-HD allows to significantly detect latent geometry variations, but not the others HyperMap-based approaches. On one side this suggests that the HyperMap approaches generally offer a lower discriminative power, but on the other side confirms the latent geometry network modifications occurring in the 2D hyperbolic space because of the pathological condition.

Taken together our results suggest that, although the human structural brain networks are weakly hyperbolic (Table 4), the coalescent embedding algorithms still offer a powerful tool for revealing the latent brain geometry. We hope that these findings will represent a convincing starting point to bridge the gap between brain networks topology and latent geometry. Finally, we believe that the introduced methodology of connectomic investigation will open a new scenario for analysing brain disorders. The 2D geometrical space for structural brain connectome representation could be used for diagnostic and prognostic purpose and for therapeutic treatment evaluation, with possible impact across many domains of neurology and psychiatry.

*Methods*

*Brain networks datasets*

The coalescent embedding methods have been tested on 4 structural human brain networks datasets.

*Dataset I:* The first healthy controls dataset was taken from a study on structural network connectivity in Parkinson's Disease patients [14], from which we included tractography-based networks of 30 healthy controls. For description and details of the construction of the structural connectivity matrices we refer to previous work of [14]. In [14] a thresholding procedure has been applied to retain only the pairwise connections with more than 3 streamlines. Each edge of the final matrices represented the product of the thresholded number of streamlines (NOS) and mean fractional anisotropy (FA) normalized by dividing each element by the maximum value of the matrix (NOS x FA / max value of the matrix).

*Dataset II:* The second dataset included 115 tractography-based connectivity matrices of healthy controls, which have been provided by the UCLA Brain Research Institute (Los Angeles, CA, United States). The strength of the connectomes refers to the region-to-region NOS. Further details on MRI data acquisition, preprocessing and processing are available in Suppl. Information.

*Dataset III:* The third healthy controls dataset, including the tractography-based connectivity matrices of 486 healthy subjects, was constructed from the T1 and diffusion weighted imaging data of the Human Connectome Project (WU-Minn HCP Data - 1200 Subjects release) [17,18]. Individual connectomes have been provided by the Dutch Connectome Lab, Utrecht, Netherlands and have been created following a procedure as explained in [19,20]. The streamlines distance (SD) between each node has been considered as the weight of the connectomes in order to reconstruct a distance network.

*Dataset IV:* this dataset is a patient-control dataset including de novo drug naïve Parkinson's Disease patients. Data was acquired on a 3T Philips Achieva clinical scanner at the IRCCS Centro Neurolesi "Bonino Pulejo", Messina, Italy. A detailed description of enrolled participants, magnetic resonance imaging data acquisition, preprocessing procedure, tractography and connectome reconstruction can be found in Suppl. Information. In the connectivity matrices, each edge represents the connectivity strength measured by NOS as provided by Constrained Spherical Deconvolution-based tractography [21,22].

Table 4 summarizes some characteristics of the brain connectomes used in the present paper.

*Evaluation of the anatomical arrangement*

In order to evaluate the latent geometry of brain networks, each network node was assigned to different anatomical classes according to three anatomical arrangements. The first anatomical arrangement refers to the left and the right hemisphere, thus each node was classified as belonging to the left or right side according to their native space location. The second one considers an anterior-central-posterior anatomical arrangement, where the central community includes the brain areas belonging to the parietal lobe and surrounding the central Rolandic fissure. Finally, for the last anatomical arrangement each cortical area has been annotated according to the brain lobes definition. If a brain region did not suit to be allocated in any anatomical class, it remained not assigned and was only considered while performing the embedding, not during the evaluation (i.e. in *Dataset II*, the node *lateral occipito-temporal sulcus* was not assigned to any anatomical lobe since it marks the border between the inferior occipital gyrus and the posterior part of the temporal lobe). The same procedure has been applied for the nodes representing the basal ganglia and the thalamus.

For each dataset and every anatomical arrangement, the evaluation has been performed according to the following procedure:

1. Given a set of connectivity matrices, an average connectivity matrix is generated using either the mean or the median operator.
2. The accuracy of the coalescent embedding techniques can be improved if the network links are weighted using values that suggest the connectivity geometry[12,13]. Since the weights in the given connectivity matrices can indicate either connection strength (i.e. in *Datasets I* and *II*) or distances between the adjacent nodes (i.e. *Dataset III*), in the first case the values need to be reversed. The assumption is that the higher the strength the higher the similarity between the adjacent nodes, therefore the lower their distance. For every edge $(i,j)$ the weight is reversed according to the following formula:
$$x'(i,j) = |x(i,j) - x_{max} - x_{min}|$$
Where $x$ is the average connectivity matrix computed at the previous point, $x'$ is the reversed average connectivity matrix, $x_{max}$ and $x_{min}$ are respectively the maximum and minimum edge weights in $x$.

    The average connectivity matrix (reversed or not according to the previous point) is embedded in the two-dimensional hyperbolic space using the coalescent embedding techniques (MCE, ncMCE, ISO, ncISO and LE), which give as output the polar

coordinates $(r, \theta)$ of the nodes in the hyperbolic disk (for details on the methods please refer to the original publication[12,13]).

3. The arrangement of the nodes over the angular coordinate space is compared to the annotated anatomical arrangement in order to evaluate the extent to which nodes belonging to the same anatomical class are close to each other in the angular coordinate (similarity) space. For this purpose, we designed two measures of circular separation. The procedure to compute them takes in input the angular coordinates of the nodes ($\theta$) and the annotated classes, and gives as output the scores in the range [0, 1]. A value 1 indicates that all the classes are perfectly separated over the circumference, with all the nodes of the same class arranged in circular sequence without interruptions. The more the classes are mixed the more the score tends to 0. The two scores that we propose are conceptually different, one is based on the average angular distance between the nodes of each class, while the other one is based on the number of wrong nodes within the extremes of each class. The scores are normalized considering the best and worst case scenarios in order to get values between 0 and 1. Suppl. Algorithm 1 describes the details of the procedure.

*Evaluation of geometrical modifications in two conditions*

In order to evaluate the geometrical modifications of the brain networks corresponding to subjects in two different conditions (i.e. two age ranges in *Dataset III* and healthy versus pathological in *Dataset IV*), we assigned a geometrical marker to every network in the dataset and then we performed tests for assessing the discrimination between the two groups.

For each connectivity matrix in the dataset, the geometrical marker has been assigned according to different methods.

1. The matrix is embedded in the 2-dimensional hyperbolic space using the coalescent embedding techniques (MCE, ncMCE, ISO, ncISO and LE as dimension reduction methods, with and without equidistant adjustment). Note that if the weights of the connectivity matrices indicate connection strength and not distances, they need to be reversed as described in the previous section. The marker is computed with two different options:

    a. Mean among all the pairwise hyperbolic distances of the nodes in the hyperbolic disk.

b. Mean among all the pairwise hyperbolic shortest paths of the nodes in the hyperbolic disk, where the hyperbolic shortest path is defined as the sum of the hyperbolic distances over the shortest path.
2. The matrix is embedded in the 3-dimensional hyperbolic space using the coalescent embedding techniques (ISO, ncISO and LE as dimension reduction methods). Note that if the weights of the connectivity matrices indicate connection strength and not distances, they need to be reversed as described in the previous section. The marker is computed with the same two different options described in 1.a and 1.b, but considering distances in the hyperbolic sphere rather than in the hyperbolic disk.
3. The matrix is embedded in the 2-dimensional hyperbolic space using the HyperMap [15] and HyperMap-CN [16] techniques. Note that in this case the matrix is treated as unweighted (for details on the methods please refer to the original publications). The marker is computed with the same two different options described in 1.a and 1.b.
4. The marker is computed as the mean of all the edge weights in the original matrix.

Once assigned for each method a geometrical marker to every connectivity matrix, three different tests are applied to the two populations of markers associated to the two groups in order to assess the ability of every method to discriminate the brain networks in the two conditions. The first test is the Wilkoxon rank-sum test, equivalent to the Mann-Whitney U-test, which is a nonparametric test for equality of population medians of two independent samples X and Y [23]. The other measures reported are the area under the receiver operating characteristic curve (AUROC or AUC for brevity) [24] and the area under the precision-recall curve (AUPR) [25], two performance scores usually adopted in classification tasks.

**Hardware and software details**

MATLAB code was used for all the methods and simulations, except for HyperMap-CN whose C code has been released by the authors at https://bitbucket.org/dk-lab/2015_code_hypermap. The simulations were carried out on a workstation under Windows 8.1 Pro with 512 GB of RAM and 2 Intel(R) Xenon(R) CPU E5-2687W v3 processors with 3.10 GHz.

**Acknowledgments**

We thank: Alexander Mestiashvili for the IT support; Claudia Matthes for the administrative support. We are grateful to Martijn van den Heuvel and the Dutch Connectome Lab to have kindly provided us the connectivity matrices used in *Dataset III*. Human neuroimaging data (for *Dataset III* of this study) was provided by the Human Connectome Project, WU-Minn

Human Connectome Project Consortium (Principal Investigators: David Van Essen and Kamil Ugurbil; Grant No. 1U54MH091657) funded by the 16 National Institutes of Health Institutes and Centers that support the National Institutes of Health Blueprint for Neuroscience Research and by the McDonnell Center for Systems Neuroscience at Washington University. C.V.C particularly thanks Trey Ideker for introducing him (during one-year visiting scholar period at UCSD in 2009) to the problem of protein-protein interactions network embedding.

*Author contributions*

CVC envisage the study, invented and designed the coalescent embedding, the algorithms and the experiments. AlbC and CVC designed the figures and wrote the article with the main aid of AM and with input and corrections from all the other authors. AM implemented and ran the codes, realized the video and performed the computational analysis with CVC help. AlbC took care of anatomical arrangement with GA, AngQ and DM help. AleC pre-processed the MRI data of *Dataset IV* and AlbC created the connectomes. SN and AldQ created and provided the connectivity matrices of *Dataset I*. EAM and JSL created and provided the connectivity matrices of *Dataset II*. AlbC and AM realized figures and tables under the CVC guidance. AlbC, AM and CVC analysed the results. CVC led, directed and supervised the study.

*Competing interests*

The authors declare no competing financial interests.

*Data and materials availability*

The MATLAB code will be available after article publication at: https://github.com/biomedical-cybernetics.

**Bibliography**

1.  Bullmore, E. & Sporns, O. Complex brain networks: graph theoretical analysis of structural and functional systems. *Nat. Rev. Neurosci.* **10,** 186–198 (2009).
2.  Bassett, D. S. & Sporns, O. Network neuroscience. *Nat. Neurosci.* **20,** 353 (2017).
3.  Hagmann, P. *et al.* Mapping the structural core of human cerebral cortex. *PLoS Biol.* **6,** 1479–1493 (2008).
4.  Gong, G. *et al.* Mapping anatomical connectivity patterns of human cerebral cortex using in vivo diffusion tensor imaging tractography. *Cereb. Cortex* **19,** 524–536

(2009).

5.  Sporns, O. & Zwi, J. D. The small world of the cerebral cortex. *Neuroinformatics* **2,** 145–162 (2004).

6.  He, Y. & Evans, A. Graph theoretical modeling of brain connectivity. *Curr Opin Neurol* **23,** 341–350 (2010).

7.  van den Heuvel, M. P. & Sporns, O. Rich-club organization of the human connectome. *J. Neurosci.* **31,** 15775–86 (2011).

8.  Bullmore, E. & Sporns, O. The economy of brain network organization. *Nat. Rev. Neurosci.* **13,** 336–349 (2012).

9.  Papadopoulos, F., Kitsak, M., Serrano, M. A., Boguna, M. & Krioukov, D. Popularity versus similarity in growing networks. *Nature* **489,** 537–540 (2012).

10. Ye, A. Q. *et al.* The intrinsic geometry of the human brain connectome. *Brain Informatics* **2,** 197–210 (2015).

11. Gulyás, A., Bíró, J. J., Kőrösi, A., Rétvári, G. & Krioukov, D. Navigable networks as Nash equilibria of navigation games. *Nat. Commun.* **6,** 7651 (2015).

12. Thomas, J. M., Muscoloni, A., Ciucci, S., Bianconi, G. & Cannistraci, C. V. Machine learning meets network science: dimensionality reduction for fast and efficient embedding of networks in the hyperbolic space. *ArXiv:1602.06522* (2016).

13. Muscoloni, A., Thomas, J. M., Ciucci, S., Bianconi, G. & Cannistraci, C. V. Machine learning meets complex networks: mapping graphs in the hyperbolic space via coalescent embedding. *(article under revision).*

14. Nigro, S. *et al.* Characterizing structural neural networks in de novo Parkinson disease patients using diffusion tensor imaging. *Hum. Brain Mapp.* **37,** 4500–4510 (2016).

15. Papadopoulos, F., Psomas, C. & Krioukov, D. Network Mapping by Replaying Hyperbolic Growth. *Networking, IEEE/ACM Trans.* **23,** 1–211 (2015).

16. Papadopoulos, F., Aldecoa, R. & Krioukov, D. Network geometry inference using common neighbors. *Phys. Rev. E - Stat. Nonlinear, Soft Matter Phys.* **92,** (2015).

17. Glasser, M. F. *et al.* The minimal preprocessing pipelines for the Human Connectome Project. *Neuroimage* **80,** 105–124 (2013).

18. Van Essen, D. C. *et al.* The Human Connectome Project: A data acquisition perspective. *Neuroimage* **62,** 2222–2231 (2012).

19. van den Heuvel, M. P., Scholtens, L. H., de Reus, M. A. & Kahn, R. S. Associated Microscale Spine Density and Macroscale Connectivity Disruptions in Schizophrenia. *Biol. Psychiatry* **80,** 293–301 (2016).

20. van den Heuvel, M. P., Scholtens, L. H., Feldman Barrett, L., Hilgetag, C. C. & de Reus, M. A. Bridging Cytoarchitectonics and Connectomics in Human Cerebral Cortex. *J. Neurosci.* **35,** 13943–13948 (2015).

21. Tournier, J. D., Calamante, F. & Connelly, A. Robust determination of the fibre orientation distribution in diffusion MRI: Non-negativity constrained super-resolved spherical deconvolution. *Neuroimage* **35,** 1459–1472 (2007).

22. Cacciola, A. *et al.* A Direct Cortico-Nigral Pathway as Revealed by Constrained Spherical Deconvolution Tractography in Humans. *Front. Hum. Neurosci.* **10,** 374 (2016).

23. Gibbons, J. D. & Chakraborti, S. *Nonparametric statistical inference*. (Chapman & Hall/Taylor & Francis, 2011).

24. Hanley, A. J. & McNeil, J. B. The Meaning and Use of the Area under a Receiver Operating Characteristic (ROC) Curve. *Radiology* **143,** 29–36 (1982).

25. Davis, J. & Goadrich, M. The Relationship Between Precision-Recall and ROC Curves. *Proc. 23rd Int. Conf. Mach. Learn. -- ICML'06* 233–240 (2006). doi:10.1145/1143844.1143874

26. Watts, D. J. & Strogatz, S. H. Collective dynamics of 'small-world' networks. *Nature* **393,** 440–2 (1998).

27. Cannistraci, C. V., Alanis-Lobato, G. & Ravasi, T. From link-prediction in brain connectomes and protein interactomes to the local-community-paradigm in complex networks. *Sci. Rep.* **3,** 1–13 (2013).

28. Clauset, A., Rohilla Shalizi, C. & J Newman, M. E. Power-Law Distributions in Empirical Data. *SIAM Rev.* **51,** 661–703 (2009).

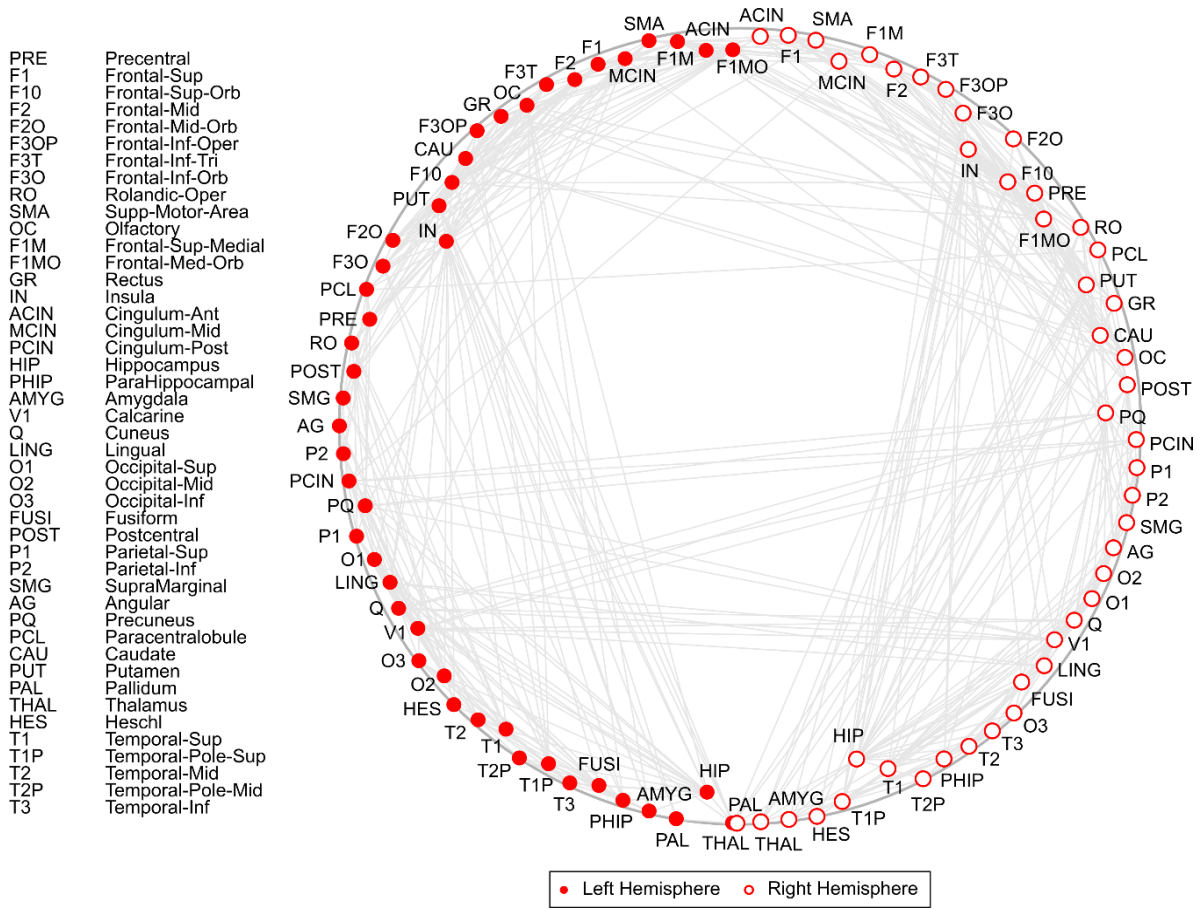

**Figure 1. Coalescent embedding on the hyperbolic disk discloses the left – right anatomical arrangement of the brain.** We show the coalescent embedding of the median matrix of *Dataset I (Nigro et al.)* using the ncMCE-EA technique. As it is possible to see from the figure and from Table 1, the structural brain networks exhibit the tendency to segregate into two sections corresponding to the left (filled circles) and right (empty circles) hemispheres. Indeed, it clearly emerges that they are perfectly separated over the disk, with all the nodes of the same hemisphere arranged in sequence without any interruption.

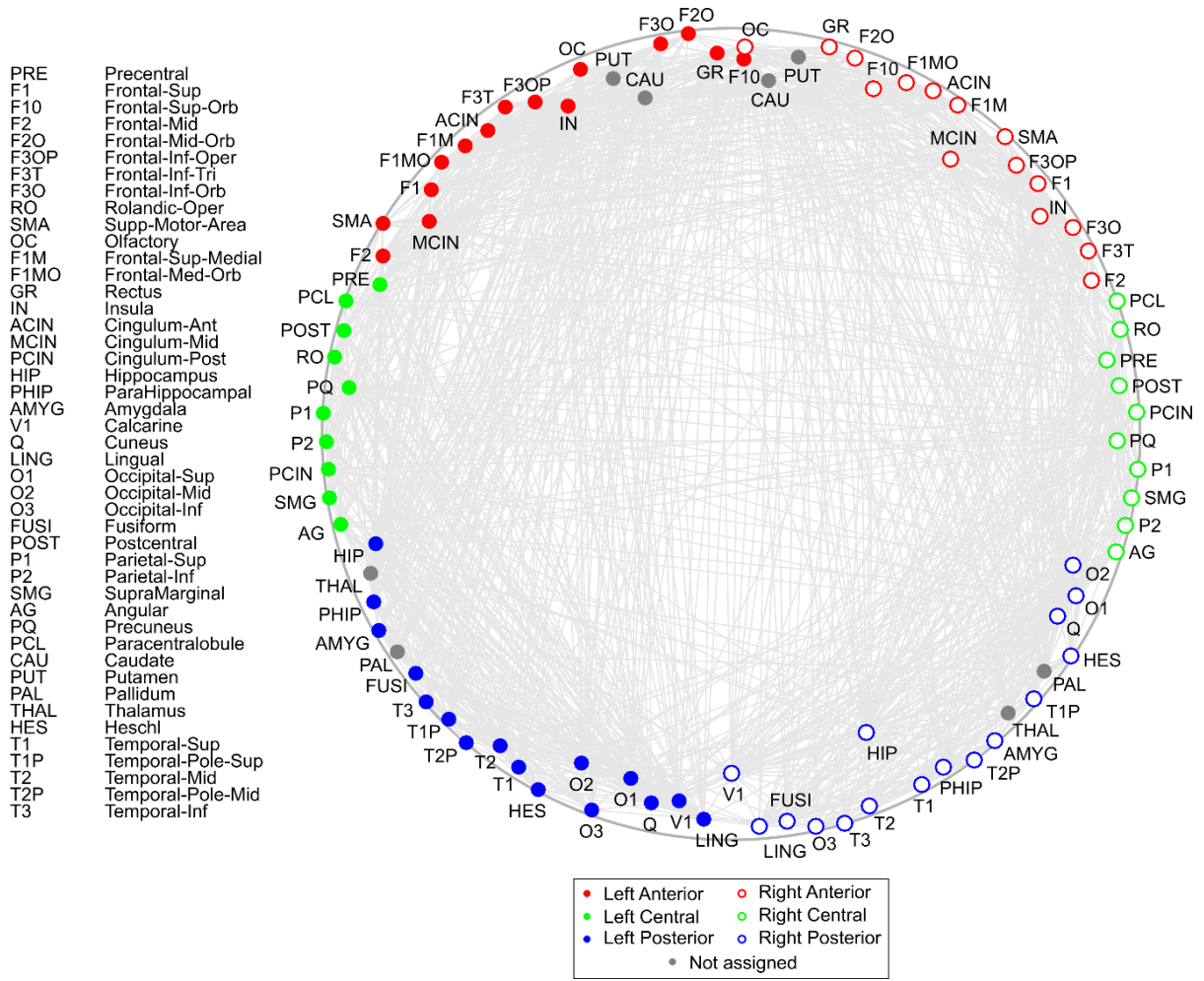

**Figure 2. Front – central – back anatomical arrangement of the brain in the hyperbolic space.** In addition to the clear segregation into two hemispheres, this figure denotes an anterior-central-posterior anatomical arrangement, where the central community includes the brain areas belonging to the parietal lobe and surrounding the central Rolandic fissure. In this case, the coalescent embedding is performed using ncMCE-EA on the mean matrix of the connectomes of *Dataset I (Nigro et al.*[14]*)*. All the groups are perfectly separated over the hyperbolic disk, apart from the filled grey nodes, which were not considered when computing the circular separation scores, since they did not suit to be included in any anatomical class due to their deep subcortical location.

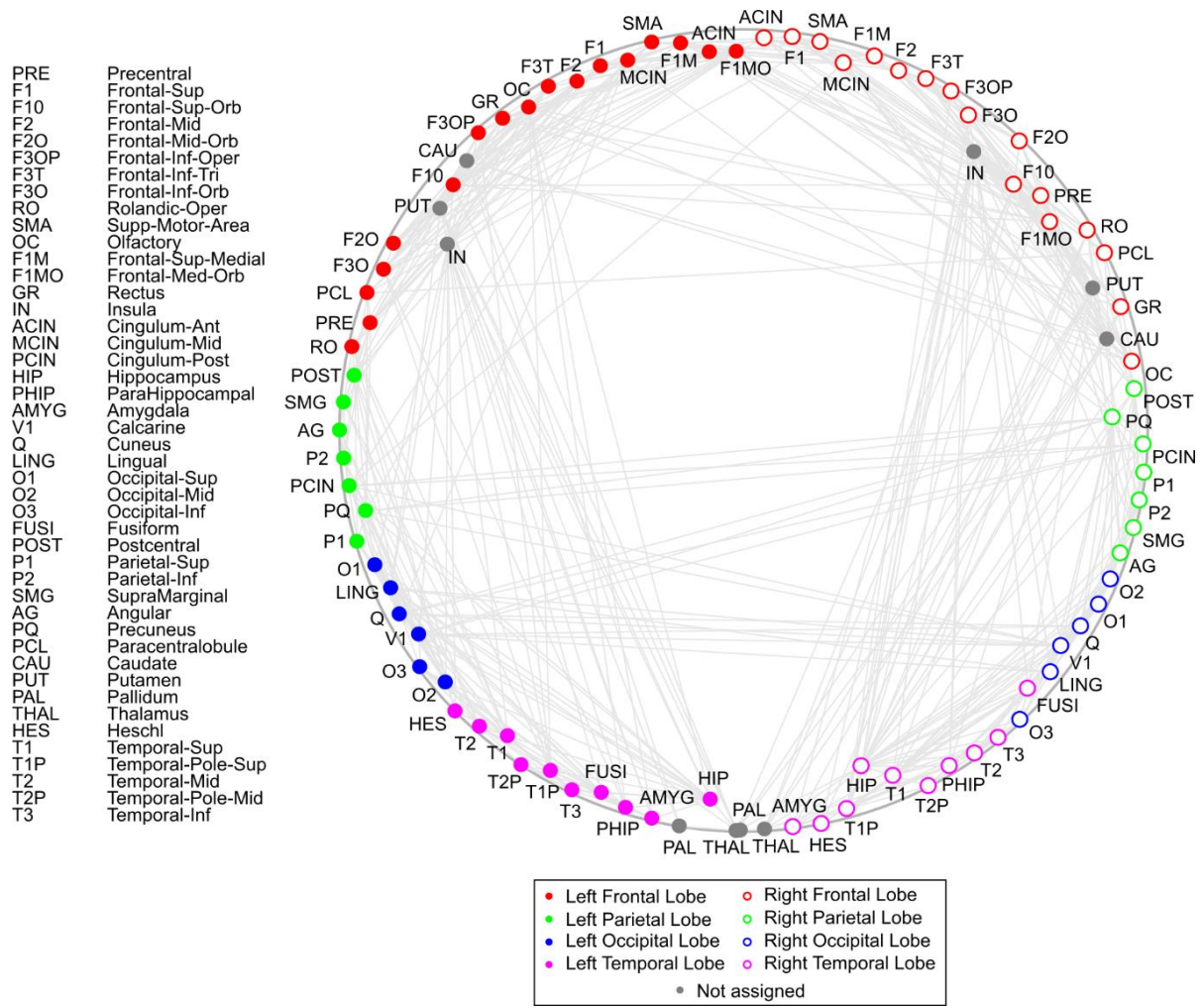

**Figure 3. Brain lobes anatomical arrangement in the hyperbolic space.** By looking only at the mere topology of the brain networks, the coalescent embedding using ncMCE-EA on the median matrix of the connectomes of *Dataset I (Nigro et al.[14])* reveals that the latent geometry of the brain strictly resembles the real anatomy of the brain lobes. Also in this case, the filled circles indicate the nodes belonging to the left hemisphere whereas the empty the ones belonging to the right hemisphere. The Frontal, Parietal, Temporal and Occipital brain lobes are clearly separated on the hyperbolic disk. The Insular Lobe has not been considered since the brain parcellation provided only one node per hemisphere (IN, grey filled circle, Not assigned). The other filled grey nodes, which have not been considered when computing the circular separation scores, represent grey matter structures placed in the deep white matter and therefore they did not suit to be included in any brain lobe. Note that the circular separation scores, reported in Table 1 for this figure, have value 1 because they are rounded, in reality there is only one mistake in the bottom-right region of the disk where one purple and one blue are swapped.

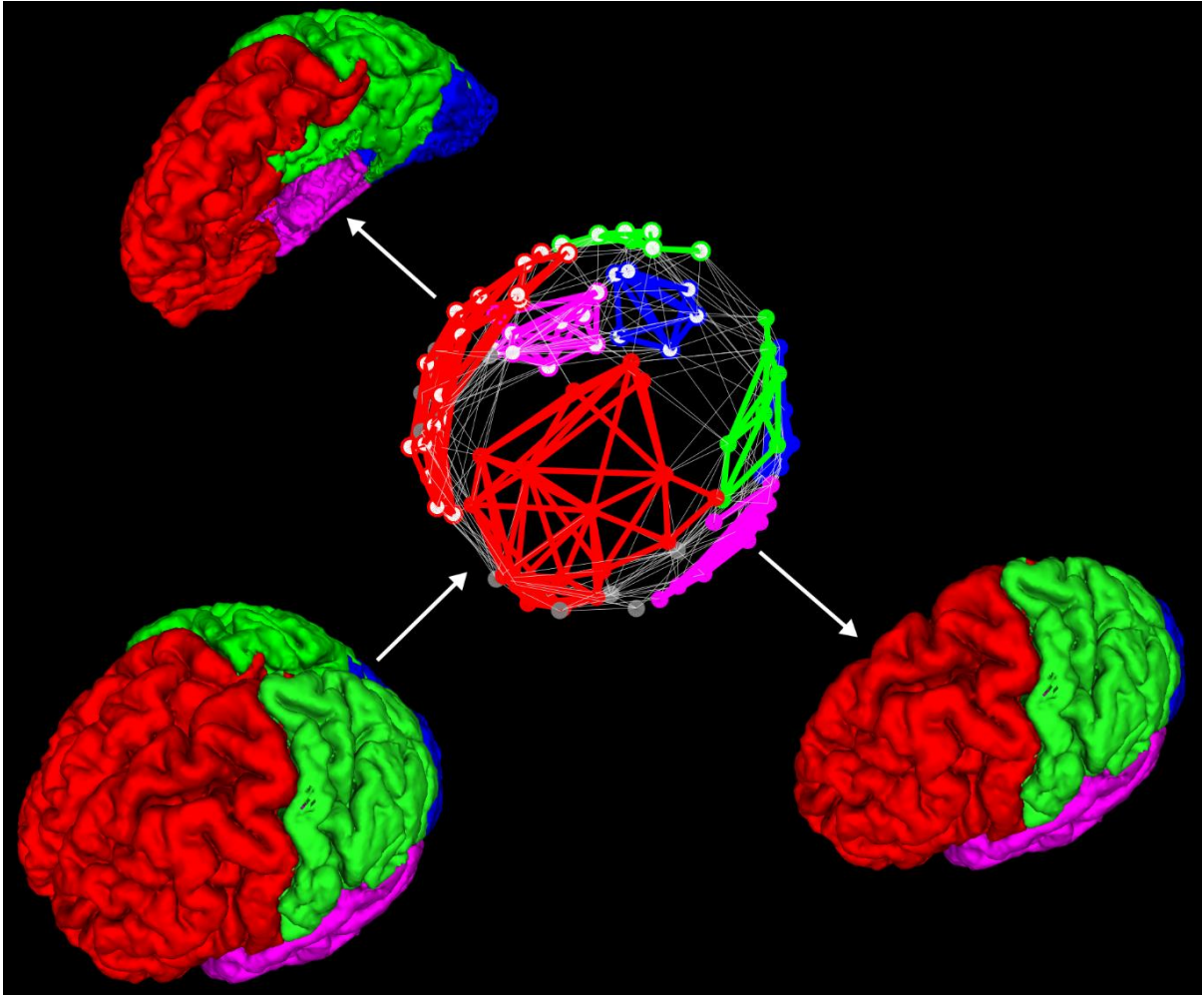

**Figure 4. ISO-3D coalescent embedding of structural brain networks highlights the presence of brain lobes anatomical arrangement.** The median connectivity matrix of 30 healthy controls (HC) of *Dataset I* has been mapped in the 3D hyperbolic space using the coalescent embedding ISO technique. The figure shows, in a superior-anterior-lateral view, the 3D geometry of the brain emerging from the embedding in the hyperbolic sphere. The colours-filled circles represent the nodes of the left hemisphere, whereas the white-filled ones represent the brain structures of the right hemisphere. Each node has been labelled according to its real anatomical localization in the different brain lobes. The red colour indicates the Frontal Lobe, the green colour the Parietal Lobe, the magenta colour the Temporal Lobe, the blue the Occipital Lobe, whereas the grey colour characterizes the nodes that have not been assigned to any lobe, since they represent grey matter structures placed in the deep white matter. It is worthy to note that the brain network geometry resembles almost perfectly the real brain anatomy, as evident from the 3D representation of a real brain. The whole brain placed anteriorly to the reconstructed network has been split into the left and right hemispheres in order to show the Right Temporal (magenta) and Occipital (blue) Lobes and to make even more visible the close relation between the latent geometry of the brain and the brain anatomy itself. Furthermore, another interesting finding is that we were able to reconstruct such latent geometry unsupervisedly starting from the mere topology of the network.

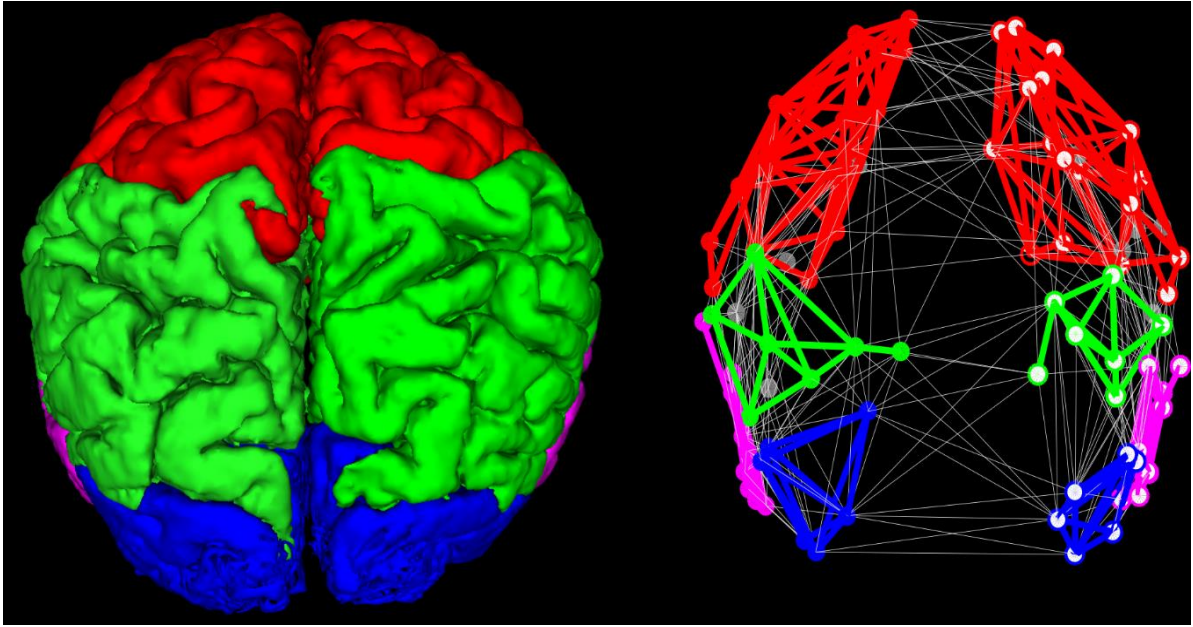

**Figure 5. ncISO-3D coalescent embedding of structural brain networks highlights the presence of brain lobes anatomical arrangement.** The median connectivity matrix of 30 healthy controls (HC) of *Dataset I* has been mapped in the 3D hyperbolic space using the coalescent embedding ncISO technique. The figure shows, in a posterior view, the 3D geometry of the brain emerging from the embedding in the hyperbolic sphere. The colours-filled circles represent the nodes of the left hemisphere, whereas the white-filled ones represent the brain structures of the right hemisphere. Each node has been labelled according to its real anatomical localization in the different brain lobes. The red colour indicates the Frontal Lobe, the green colour the Parietal Lobe, the magenta colour the Temporal Lobe, the blue the Occipital Lobe, whereas the grey colour characterizes the nodes that have not been assigned to any lobe, since they represent grey matter structures placed in the deep white matter.

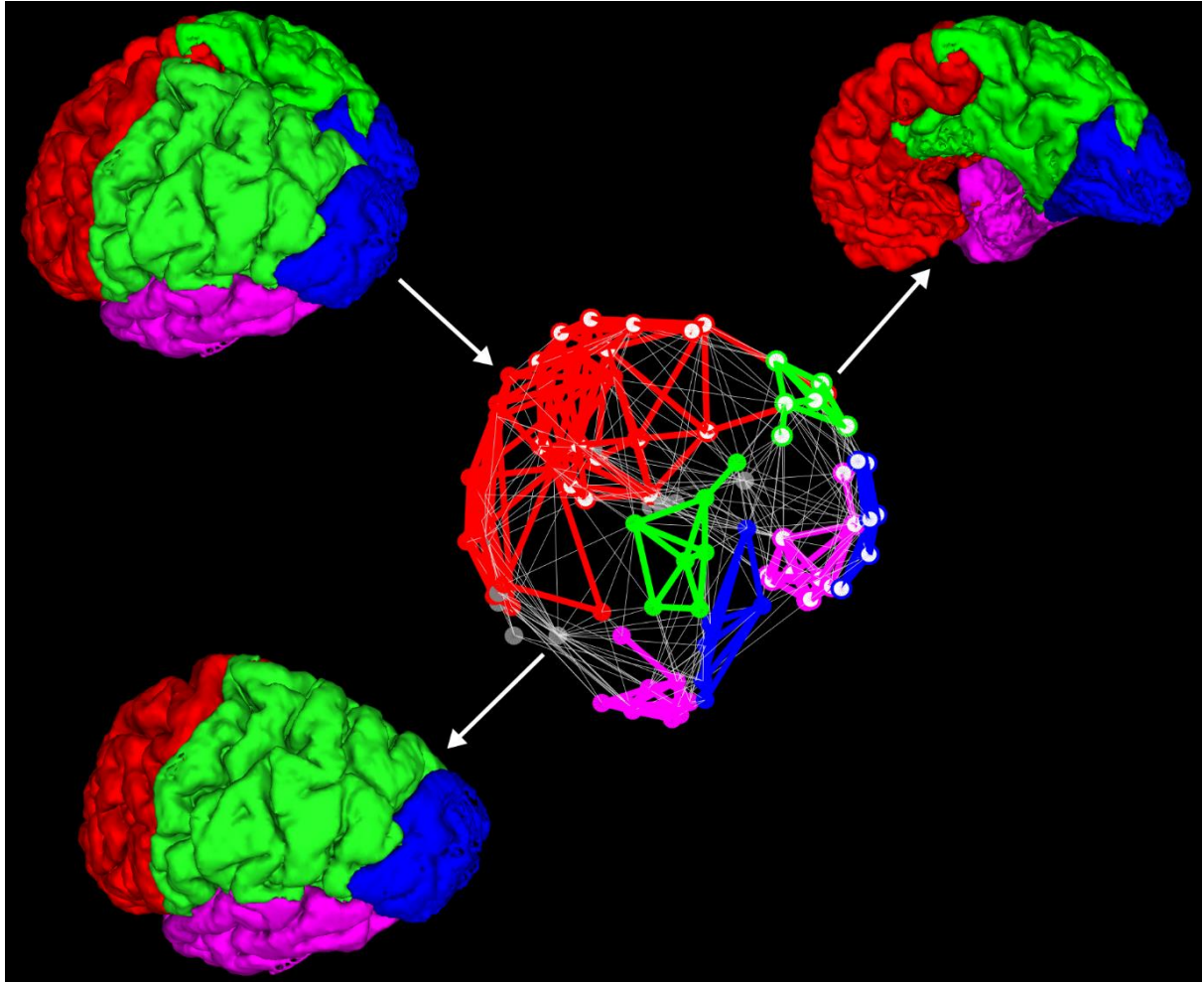

**Figure 6. LE-3D coalescent embedding of structural brain networks highlights the presence of brain lobes anatomical arrangement.** The median connectivity matrix of 30 healthy controls (HC) of *Dataset I* has been mapped in the 3D hyperbolic space using the coalescent embedding LE technique. The figure shows, in a superior-posterior-lateral view, the 3D geometry of the brain emerging from the embedding in the hyperbolic sphere. The colours-filled circles represent the nodes of the left hemisphere, whereas the white-filled ones represent the brain structures of the right hemisphere. Each node has been labelled according to its real anatomical localization in the different brain lobes. The red colour indicates the Frontal Lobe, the green colour the Parietal Lobe, the magenta colour the Temporal Lobe, the blue the Occipital Lobe, whereas the grey colour characterizes the nodes that have not been assigned to any lobe, since they represent grey matter structures placed in the deep white matter. The whole brain placed anteriorly to the reconstructed network has been split into the left and right hemispheres in order to show the Right Temporal (magenta) and Occipital (blue) Lobes and to make even more visible the close relation between the latent geometry of the brain and the brain anatomy itself. Note that we were able to reconstruct such latent geometry unsupervisedly starting from the mere topology of the network and both ISO (Figure 4), ncISO (Figure 5) and LE (in this figure) achieved the task with high accuracy.

| | | NIGRO | | | LABUS | | | VAN DEN HEUVEL | |
|---|---|---|---|---|---|---|---|---|---|
| | | score-d | score-w | | score-d | score-w | | score-d | score-w |
| **LEFT - RIGHT** | **ISO-mean** | 1.00 | 1.00 | **LE-median** | 0.99 | 0.95 | **ISO-median** | 1.00 | 1.00 |
| | **ncISO-mean** | 1.00 | 1.00 | **ISO-mean** | 0.99 | 0.95 | **ncISO-median** | 1.00 | 1.00 |
| | **MCE-mean** | 1.00 | 1.00 | ncISO-median | 0.99 | 0.94 | **LE-median** | 1.00 | 1.00 |
| | **ncMCE-mean** | 1.00 | 1.00 | ncISO-mean | 0.98 | 0.95 | ncMCE-median | 0.71 | 0.54 |
| | **LE-mean** | 1.00 | 1.00 | ISO-median | 0.98 | 0.93 | MCE-median | 0.22 | 0.31 |
| | **ISO-median** | 1.00 | 1.00 | LE-mean | 0.98 | 0.94 | MCE-mean | 0.07 | 0.18 |
| | **ncISO-median** | 1.00 | 1.00 | ncMCE-median | 0.32 | 0.31 | LE-mean | 0.07 | 0.19 |
| | **MCE-median** | 1.00 | 1.00 | MCE-median | 0.31 | 0.36 | ncMCE-mean | 0.02 | 0.16 |
| | **ncMCE-median** | 1.00 | 1.00 | MCE-mean | 0.07 | 0.22 | ncISO-mean | 0.01 | 0.11 |
| | **LE-median** | 1.00 | 1.00 | ncMCE-mean | 0.06 | 0.31 | ISO-mean | 0.01 | 0.15 |
| | | score-d | score-w | | score-d | score-w | | score-d | score-w |
| **FRONT - CENTRAL - BACK** | **ncMCE-mean** | 1.00 | 1.00 | **ncISO-median** | 0.73 | 0.76 | **ncISO-median** | 0.81 | 0.85 |
| | ncMCE-median | 0.98 | 0.98 | ISO-median | 0.72 | 0.77 | **LE-median** | 0.81 | 0.87 |
| | MCE-mean | 0.96 | 0.95 | LE-median | 0.51 | 0.63 | ISO-median | 0.80 | 0.85 |
| | MCE-median | 0.92 | 0.94 | MCE-median | 0.45 | 0.53 | ncMCE-median | 0.66 | 0.70 |
| | LE-mean | 0.87 | 0.89 | LE-mean | 0.42 | 0.60 | LE-mean | 0.52 | 0.57 |
| | ncISO-mean | 0.82 | 0.85 | ISO-mean | 0.42 | 0.61 | MCE-median | 0.40 | 0.54 |
| | ncISO-median | 0.81 | 0.86 | ncISO-mean | 0.42 | 0.61 | ncISO-mean | 0.14 | 0.32 |
| | ISO-median | 0.81 | 0.86 | ncMCE-median | 0.35 | 0.37 | ISO-mean | 0.11 | 0.30 |
| | LE-median | 0.81 | 0.86 | ncMCE-mean | 0.32 | 0.50 | MCE-mean | 0.06 | 0.27 |
| | ISO-mean | 0.81 | 0.83 | MCE-mean | 0.30 | 0.47 | ncMCE-mean | 0.06 | 0.24 |
| | | score-d | score-w | | score-d | score-w | | score-d | score-w |
| **BRAIN LOBES** | **ncMCE-median** | 1.00 | 1.00 | **ISO-median** | 0.75 | 0.80 | **ncISO-median** | 0.84 | 0.85 |
| | ncMCE-mean | 0.96 | 0.97 | ncISO-median | 0.74 | 0.79 | ISO-median | 0.83 | 0.85 |
| | MCE-median | 0.96 | 0.96 | LE-median | 0.54 | 0.65 | LE-median | 0.83 | 0.87 |
| | MCE-mean | 0.93 | 0.94 | ISO-mean | 0.52 | 0.68 | ncMCE-median | 0.65 | 0.67 |
| | LE-median | 0.89 | 0.92 | ncISO-mean | 0.52 | 0.69 | LE-mean | 0.62 | 0.64 |
| | ncISO-median | 0.88 | 0.91 | LE-mean | 0.50 | 0.67 | MCE-median | 0.34 | 0.48 |
| | ISO-median | 0.87 | 0.91 | MCE-median | 0.50 | 0.58 | ncISO-mean | 0.20 | 0.41 |
| | LE-mean | 0.85 | 0.85 | ncMCE-median | 0.49 | 0.58 | ISO-mean | 0.16 | 0.31 |
| | ncISO-mean | 0.83 | 0.85 | ncMCE-mean | 0.30 | 0.49 | MCE-mean | 0.09 | 0.25 |
| | ISO-mean | 0.82 | 0.83 | MCE-mean | 0.29 | 0.50 | ncMCE-mean | 0.08 | 0.23 |

**Table 1. Circular separation scores for each anatomical arrangement and for the three datasets analysed.** The table reports for each anatomical arrangement and for each dataset, the two circular separation scores for all the coalescent embedding techniques ncMCE, MCE, ncISO, ISO and LE. Note that the equidistant adjustment (EA) does not affect the circular separation, therefore the scores are not reported, since identical to the ones of the non-EA variants. The scores evaluate the extent to which nodes belonging to the same annotated anatomical class are close to each other in the angular coordinate space of the hyperbolic embedding. They assume values in the range [0, 1]. A value 1 indicates that all the classes are perfectly separated over the circumference, with all the nodes of the same class arranged in circular sequence without interruptions. The more the classes are mixed the more the score tends to 0 (for details on the scores please refer to Suppl. Algorithm 1). The suffix mean or median indicates the method according to which the average connectivity matrix to embed has been obtained. Note that the scores reported have been rounded at the second digit and the best method according to the circular separation *score-d* is highlighted in bold.

|  | mean marker (age 21-25) | mean marker (age 31-35) | MW p-value | AUC | AUPR |
|---|---|---|---|---|---|
| **ISO-HD** | 14.8 | 14.9 | **0.005** | 0.61 | 0.66 |
| **ncISO-HD** | 14.8 | 14.9 | **0.007** | 0.61 | 0.66 |
| **ncISO-HSP** | 20.1 | 20.5 | **0.008** | 0.61 | 0.65 |
| **ISO-HSP** | 20.1 | 20.5 | **0.011** | 0.60 | 0.66 |
| **ncMCE-HD** | 14.2 | 14.4 | **0.017** | 0.60 | 0.67 |
| **ncMCE-HSP** | 18.7 | 19.1 | **0.017** | 0.60 | 0.68 |
| **MCE-HD** | 14.6 | 14.7 | **0.022** | 0.59 | 0.65 |
| **ncISO-EA-HD** | 15.0 | 15.1 | **0.022** | 0.59 | 0.68 |
| **ISO-EA-HD** | 15.0 | 15.1 | **0.022** | 0.59 | 0.68 |
| **ncMCE-EA-HD** | 15.0 | 15.1 | **0.022** | 0.59 | 0.68 |
| **MCE-EA-HD** | 15.0 | 15.1 | **0.022** | 0.59 | 0.68 |
| **LE-EA-HD** | 15.0 | 15.1 | **0.022** | 0.59 | 0.68 |
| **LE-HD** | 14.9 | 15.0 | **0.039** | 0.58 | 0.68 |
| **MCE-HSP** | 19.9 | 20.3 | **0.042** | 0.58 | 0.66 |
| **ncISO-EA-HSP** | 21.5 | 21.8 | **0.044** | 0.58 | 0.67 |
| **ISO-EA-HSP** | 21.5 | 21.8 | **0.050** | 0.58 | 0.67 |
| **MCE-EA-HSP** | 21.6 | 21.9 | 0.074 | 0.57 | 0.68 |
| **LE-EA-HSP** | 21.3 | 21.6 | 0.078 | 0.57 | 0.68 |
| **ncMCE-EA-HSP** | 21.6 | 21.9 | 0.080 | 0.57 | 0.68 |
| **LE-HSP** | 20.4 | 20.6 | 0.160 | 0.56 | 0.69 |
| **weights** | 80.4 | 79.9 | 0.311 | 0.54 | 0.72 |

**Table 2. Geometrical modifications of the brain in two different age ranges.** Two subsamples of connectivity matrices of healthy controls (22-25 and 31-35 age-range, respectively) included *in Dataset III (van den Heuvel)*, have been mapped using the coalescent embedding algorithms. The networks represent the Streamlines Distance (SD) between pairwise nodes as provided by Generalized Q-sampling imaging (GQI) tractography. The table reports for each method (both with the EA and non-EA variants), the mean marker of the two groups, the p-value of the Mann-Whitney (MW) test, the area under the receiver operating characteristic curve (AUC) and the area under precision-recall curve (AUPR). The suffix HD or HSP indicates if the marker in the hyperbolic space is the average hyperbolic distance or hyperbolic shortest path. For reference, also the marker computed as the average edge weight in the original network is reported. Note that significant p-values are highlighted in bold, considering a confidence level of 0.05. As emerging from the table, almost all the methods uncover a significantly different geometry underlying the human structural brain networks in two age range-specific groups.

|  | mean marker (HC) | mean marker (PD) | MW p-value | AUC | AUPR |
|---|---|---|---|---|---|
| **MCE-HSP** | 14.7 | 15.7 | **0.006** | 0.87 | 0.82 |
| **ncMCE-HSP** | 14.3 | 15.2 | **0.014** | 0.83 | 0.79 |
| **MCE-HD** | 14.4 | 15.3 | **0.017** | 0.82 | 0.77 |
| **ncMCE-HD** | 14.1 | 14.9 | **0.017** | 0.82 | 0.77 |
| **LE-HSP** | 15.6 | 16.3 | **0.017** | 0.82 | 0.78 |
| **ncISO-EA-HSP** | 15.9 | 16.7 | **0.021** | 0.81 | 0.78 |
| **ncMCE-EA-HSP** | 15.9 | 16.7 | **0.021** | 0.81 | 0.78 |
| **ISO-EA-HSP** | 15.9 | 16.7 | **0.026** | 0.80 | 0.77 |
| **ncISO-HSP** | 15.5 | 16.2 | **0.026** | 0.80 | 0.76 |
| **MCE-EA-HSP** | 15.9 | 16.7 | **0.026** | 0.80 | 0.77 |
| **LE-EA-HSP** | 15.9 | 16.7 | **0.026** | 0.80 | 0.77 |
| **LE-HD** | 15.0 | 15.6 | **0.038** | 0.78 | 0.72 |
| **ISO-HD** | 14.8 | 15.4 | **0.045** | 0.77 | 0.71 |
| **ISO-HSP** | 15.4 | 16.2 | **0.045** | 0.77 | 0.73 |
| **ncISO-HD** | 14.9 | 15.5 | 0.121 | 0.71 | 0.64 |
| **ncISO-EA-HD** | 15.2 | 15.7 | 0.121 | 0.71 | 0.69 |
| **LE-EA-HD** | 15.2 | 15.7 | 0.121 | 0.71 | 0.69 |
| **ISO-EA-HD** | 15.2 | 15.7 | 0.186 | 0.68 | 0.66 |
| **MCE-EA-HD** | 15.2 | 15.7 | 0.186 | 0.68 | 0.66 |
| **ncMCE-EA-HD** | 15.2 | 15.7 | 0.186 | 0.68 | 0.66 |
| **weights** | 222.0 | 215.8 | 0.307 | 0.64 | 0.64 |

**Table 3. Geometrical modifications of the brain in *de novo* drug naïve Parkinson's Disease (PD) patients compared to Healthy Controls (HC).** The connectomes of 10 PD patients and 10 age- and sex-matched HC have been mapped using the coalescent embedding algorithms. The networks represent the Number of Streamlines (NOS) between pairwise regions as provided by Constrained Spherical Deconvolution (CSD) tractography. The table reports for each method (both with the EA and non-EA variants), the mean marker of the two groups, the p-value of the Mann-Whitney (MW) test, the area under the receiver operating characteristic curve (AUC) and the area under precision-recall curve (AUPR). The suffix HD or HSP indicates if the marker in the hyperbolic space is the average hyperbolic distance or hyperbolic shortest path. For reference, also the marker computed as the average edge weight in the original network is reported. Note that significant p-values are highlighted in bold, considering a confidence level of 0.05. As emerging from the table, almost all the methods uncover a significantly different geometry underlying the human structural brain networks in *de novo* drug naïve PD patients.

|  | **Nigro** | **Labus** | **Van den Heuvel** | **Cacciola** |
|---|---|---|---|---|
| **Condition** | healthy controls 30 | healthy controls 115 | healthy controls 486 | controls / pathological 10 / 10 |
| **N** | 90 | 165 | 82 | 84 |
| **E** | 561 | 1556 | 1164 | 3102 |
| **avg degree** | 12.5 | 18.9 | 28.4 | 73.9 |
| **density** | 0.14 | 0.12 | 0.35 | 0.89 |
| **clustering** | 0.54 | 0.53 | 0.65 | 0.92 |
| **char path** | 2.45 | 2.30 | 1.69 | 1.11 |
| **LCP-corr** | 0.95 | 0.96 | 0.98 | 1.00 |
| **powerlaw-$\gamma$** | 7.03 | 4.79 | 4.32 | 7.74 |
| **powerlaw-p** | 0.00 | 0.06 | 0.01 | 0.00 |
| **Technique** | FACT tractography | FACT tractography | GQI tractography | CSD probabilistic tractography |
| **Weight** | NOS(thr3) x FA / max | NOS | SD | NOS |

**Table 4. Summary of brain networks characteristics.** The table reports for each dataset the number of healthy controls and pathological subjects. For each network, several statistics have been computed and the mean over the dataset is reported. Number of nodes N, number of edges E, average node degree, network density, average clustering coefficient, computed for each node as the number of links between its neighbours over the number of possible links [26], characteristic path length of the network [26]. LCP-corr is the Local Community Paradigm correlation [27], representing the correlation between the number of common neighbours and the number of links between them, looking at each pair of connected nodes in the network. Powerlaw-$\gamma$ is the exponent of the power-law distribution estimated from the observed degree distribution of the network using the maximum likelihood procedure described in [28], whereas powerlaw-p is the p-value for the estimated power-law fit to the data, considered significant when > 0.10 as suggested in the original publication [28]. The tractographic algorithms (FACT = Fiber assignment by continuous tracking; GQI = Generalized Q-sampling imaging; CSD = Constrained Spherical Deconvolution) used to create the connectomes and their weights (NOS(thr3) x FA / max = Product between Number of Streamlines (cut off = 3) and Fractional Anisotropy (FA) normalized by the maximum value of the matrix; NOS = Number of Streamlines; SD = Streamlines Distance).

# Supplementary Information

**Methods**

*Dataset III (Labus, UCLA, Los Angeles, CA, United States)*
*MRI Acquisition, quality control, preprocessing and processing.*
Whole brain structural (diffusion tensor imaging, DTI) data was acquired from 115 healthy subjects (58Males, 57Females) using a 3.0T MRI scanner (Siemens Trio; Siemens, Erlangen, Germany). Detailed information on the standardized acquisition protocols, quality control measures, and image preprocessing are provided in previously published studies [1–5].

*Structural gray-matter.* Structural T1-image segmentation and regional parcellation were conducted using FreeSurfer [6,7] following the nomenclature described in Destrieux et al. [6] and the Harvard-Oxford subcortical Atlas. This parcellation results in the labeling of 165 regions, 74 bilateral cortical structures, 7 subcortical structures, the brainstem, and the cerebellum.

*Anatomical network construction.* Regional parcellation and tractography results were combined to produce a weighted, undirected connectivity matrix. White matter connectivity for each subject was estimated between the 165 brain regions using DTI fiber tractography [3], performed via the Fiber Assignment by Continuous Tracking (FACT) algorithm [8] using TrackVis (http://trackvis.org). The final estimate of white matter connectivity between each of the brain regions was determined based on the number of fiber tracts intersecting each region. Weights of the connections were then expressed as the absolute fiber count divided by the individual volumes of the two interconnected regions [3].

*Dataset IV (Cacciola, IRCCS Centro Neurolesi "Bonino Pulejo", Messina, Italy)*
*Participants*
Ten healthy subjects (mean age 58.5±5.87; 6 Males, 4 Females) and ten *de novo* drug-naïve Parkinson's Disease (PD) patients (mean age 62.9±5.56 years; 5 Males, 5 Females) were recruited from the Movement Disorders Clinic of IRCCS Centro Neurolesi of Messina. PD diagnosis was made by a board-certified movement disorders specialist, using the United Kingdom Parkinson's Disease Society Brain Bank Criteria as a guide [9].

Exclusion criteria covered a broad range of conditions such as: i) parkinsonism due to antipsychotics or other drugs; ii) suspected dementia with Lewy bodies; iii) transient loss of consciousness; iv) delirium; v) confusion; vi) amnestic disorder; vii) neuropsychiatric diseases;

viii) cerebral vascular lesions; ix) post traumatic brain injury on MRI; x) patients in treatment with antiparkinsonian drugs; xi) Patients with progressive supranuclear palsy and multiple system atrophy.

The Unified Idiopathic Parkinson's Disease Rating Scale (UPDRS) [10] and the Hoehn and Yahr staging [11] have been used to evaluate disease severity. All patients were screened for cognitive impairment and depression via the Montreal Cognitive Assessment (MoCA) [12] and the Beck Depression Inventory (BDI-II) [13], respectively. Demographic and clinical data (mean±SD) of PD patients and healthy controls are reported in Suppl. Table 4.

All individuals read and signed informed consent before examinations. The entire study protocol was in accordance to the Declaration of Helsinki and was approved by the Institutional Review Board of IRCCS Bonino Pulejo - Messina - Italy (Scientific Institute for Research, Hospitalization and Health Care).

*MRI Acquisition, quality control, preprocessing and processing.*

The following MRI sequences were acquired using a 3T Achieva Philips scanner equipped with a 32-channels SENSE head coil (Best, Netherlands):

- 3D high-resolution T1 weighted Fast Field Echo (FFE) sequence was acquired using the following parameters: TR=25 ms; TE=4.6 ms; flip angle=30°; FOV=240×240=mm2; reconstruction matrix=240x240 voxel; voxel size 1x1x1 mm; slice thickness 1 mm.
- Dual phase encoded pulsed gradient spin echo Diffusion Weighted sequences using 32 gradient diffusion directions chosen by following an electrostatic repulsion model, more 1 un-weighted b0 volume. The other sequence parameters were: diffusion weighting b-factor=1000 s/mm2; TR=11884 ms; TE=54 ms; FOV=240×240 mm2; reconstruction matrix 120x120 voxel; in-plane voxel size 2x2 mm$^2$; axial slice thickness 2 mm; no inter-slice gap.

Detailed information on image preprocessing, diffusion signal modeling and tractography are provided in previous works [14–18]. Briefly image preprocessing included: i) realignment and reorientation of individual images (both DWIs and T1) to the anterior commissure so that each part of the brain in all volumes was in the same position; ii) motion and susceptibility distortion artefacts correction using tools available within SPM8 Matlab toolbox (http://www.fil.ion.ucl.ac.uk/spm/software/spm8/); iii) co-registration of T1 images onto preprocessed diffusion images following the pipeline explained in[19], using New Segment option of SPM8 (http://www.fil.ion.ucl.ac.uk/spm/software/spm8/) as well as FLIRT and FNIRT FSL utilities (http://fsl.fmrib.ox.ac.uk/fsl/fslwiki/).

*Structural gray-matter.* For both controls and PD subjects, cortical parcellation and subcortical segmentation were performed on co-registered T1 images with the default reconstruction pipeline of Freesurfer image analysis suite by using the Desikan-Killiany atlas [20], which is documented and freely available for download online [21] (http://surfer.nmr.mgh.harvard.edu/). In this way, we obtained a total of 84 nodes, including sub-cortical GM nuclei and cerebellar hemispheres. However, FreeSurfer may lead to high variability in spatial location and extent of deep GM structures [22], thus we overcame this problem by replacing them with the more biologically accurate segmentations provided by FSL's FIRST tool [23]. The quality of parcellation and segmentation was manually checked for each subject by two of the authors (Alb. C. and Ale. C.)

*Anatomical network construction.* Diffusion signal was modeled using a modified High Angular Resolution Diffusion Imaging (HARDI) technique, namely CSD which consists in estimating, for each voxel, a fiber Orientation Distribution Function (fODF). fODF is a continuous function of the sphere which reflects the number and direction of the orientations within a given voxel and their relative weightings. Further information on CSD technique can be found in [14–18].

CSD probabilistic whole brain tractography by generating one million streamlines using WM masks previously estimated both as seed and mask ROIs. Before this step, a small dilatation to WM masks was applied in order to allow streamlines to reach our ROIs, placed in GM, for subsequent analyses. Both fODF estimation and tractography were performed by using MRtrix3 software package (http://www.mrtrix.org). For each subject, streamlines were mapped to the relevant nodes defined by the parcellation of that subject's anatomical image. In line with other studies [24,25], we considered each brain structure as a node and the inter-regional number of streamlines as edges to construct an 84x84 connectivity matrix $C_{ij}=[c_{ij}]$, as the connection strength between pairwise nodes is usually measured by the number of streamlines via which they are interconnected [26–28].

|  | mean marker (age 21-25) | mean marker (age 31-35) | MW p-value | AUC | AUPR |
|---|---|---|---|---|---|
| **ncISO3D-HD** | 15.1 | 15.3 | **0.006** | 0.61 | 0.66 |
| **ISO3D-HD** | 15.1 | 15.2 | **0.008** | 0.61 | 0.66 |
| **ncISO3D-HSP** | 21.7 | 22.1 | **0.010** | 0.60 | 0.66 |
| **ISO3D-HSP** | 21.6 | 22.0 | **0.016** | 0.60 | 0.66 |
| **LE3D-HD** | 15.3 | 15.4 | **0.024** | 0.59 | 0.68 |
| **LE3D-HSP** | 22.2 | 22.5 | 0.064 | 0.57 | 0.68 |
| **weights** | 80.4 | 79.9 | 0.311 | 0.54 | 0.72 |

**Suppl. Table 1. Geometrical modifications in the 3D embedding space of the brain in two different age ranges.** Two subsamples of connectivity matrices of healthy controls (22-25 and 31-35 age-range, respectively) included in Dataset III (van den Heuvel), have been mapped using the coalescent embedding algorithms in the 3D hyperbolic sphere. The networks represent the Streamlines Distance (SD) between pairwise nodes as provided by Generalized Q-sampling imaging (GQI) tractography. The table reports for the methods ncISO, ISO and LE the mean marker of the two groups, the p-value of the Mann-Whitney (MW) test, the area under the receiver operating characteristic curve (AUC) and the area under precision-recall curve (AUPR). The suffix HD or HSP indicates if the marker in the hyperbolic space is the average hyperbolic distance or hyperbolic shortest path. For reference, also the marker computed as the average edge weight in the original network is reported. Note that significant p-values are highlighted in bold, considering a confidence level of 0.05. The table underlines that the addition of the third dimension does not lead to a significant improvement with respect to the hyperbolic disk, but only to weak oscillations between a small increase and a little decrease of performance depending on the method.

|  | mean marker (HC) | mean marker (PD) | MW p-value | AUC | AUPR |
|---|---|---|---|---|---|
| **ISO3D-HD** | 15.2 | 15.8 | **0.021** | 0.81 | 0.77 |
| **LE3D-HD** | 15.4 | 16.0 | **0.031** | 0.79 | 0.77 |
| **ISO3D-HSP** | 16.1 | 17.0 | **0.038** | 0.78 | 0.76 |
| **ncISO3D-HSP** | 16.2 | 17.1 | **0.038** | 0.78 | 0.76 |
| **ncISO3D-HD** | 15.3 | 15.9 | **0.045** | 0.77 | 0.74 |
| **LE3D-HSP** | 16.3 | 17.1 | **0.045** | 0.77 | 0.76 |
| **weights** | 222.0 | 215.8 | 0.307 | 0.64 | 0.64 |

**Suppl. Table 2. Geometrical modifications in the 3D embedding space of the brain in *de novo* drug naïve Parkinson's Disease (PD) patients compared to Healthy Controls (HC).** The connectomes of 10 PD patients and 10 age- and sex-matched HC have been mapped using the coalescent embedding algorithms in the 3D hyperbolic sphere. The networks represent the Number of Streamlines (NOS) between pairwise regions as provided by Constrained Spherical Deconvolution (CSD) tractography. The table reports for the methods ncISO, ISO and LE the mean marker of the two groups, the p-value of the Mann-Whitney (MW) test, the area under the receiver operating characteristic curve (AUC) and the area under precision-recall curve (AUPR). The suffix HD or HSP indicates if the marker in the hyperbolic space is the average hyperbolic distance or hyperbolic shortest path. For reference, also the marker computed as the average edge weight in the original network is reported. Note that significant p-values are highlighted in bold, considering a confidence level of 0.05. As emerging from the table, all the 3D coalescent embedding methods uncover a significantly different geometry underlying the human structural brain networks in healthy versus pathological subjects. In particular all the methods, except ncISO-HSP, show a slight improvement compared to the 2D hyperbolic space (Table 3).

|  | mean marker (HC) | mean marker (PD) | MW p-value | AUC | AUPR |
| --- | --- | --- | --- | --- | --- |
| **HyperMap-HD** | 13.0 | 14.0 | **0.026** | 0.80 | 0.76 |
| **HyperMapCN-HD** | 12.4 | 13.9 | 0.076 | 0.74 | 0.66 |
| **HyperMap-HSP** | 13.7 | 14.8 | 0.076 | 0.74 | 0.69 |
| **HyperMapCN-HSP** | 12.9 | 14.5 | 0.104 | 0.72 | 0.65 |

**Suppl. Table 3. Geometrical modifications of the brain in *de novo* drug naïve Parkinson's Disease (PD) patients compared to Healthy Controls (HC) using HyperMap-based methods.** The table is equivalent to Table 3 in the main article, but reports the performance of the HyperMap-based hyperbolic embedding methods. It shows that only HyperMap-HD allows to unveil PD-related latent geometry variations, therefore the HyperMap-based approaches in general offer less discriminative power compared to the other coalescent embedding methods.

| Demographic and clinical data | PD patients (n=10) | HC (n=10) |
| --- | --- | --- |
| Gender (M/F) | 5/5 | 6/4 |
| Age (years ± SD) | 62.9 ± 5.56 | 58.5 ± 5.87 |
| UPDRS-III (mean ± SD) | 18.1± 3.63 | NA |
| H&Y (mean ± SD) | 1.5 ± 0,53 | NA |
| MoCA (mean ± SD) | 27.8 ± 0.79 | NA |
| BDI-II (mean ± SD) | 9.7 ± 2.31 | NA |
| Treatment | NA | NA |

**Suppl. Table 4. Demographic and clinical data (mean ± SD) of *de novo* drug naïve Parkinson's disease (PD) patients and sex- and age-matched healthy controls included in *Dataset IV*.**

Legend: UPDRS - III = Unified Parkinson's Disease Rating Scale – score III; H&Y = Hoehn and Yahr staging; MoCA = Montreal Cognitive Assessment; BDI-II = Beck Depression Inventory - II; NA = Not Applicable. It is noteworthy that the selected patients are *de novo* and untreated, thus removing possible effects of disease duration and therapy.

**Suppl. Algorithm 1.** Procedure to compute the circular separation measures *score-d* and *score-w*.

---

INPUT: $\theta_{1...N}$, $C_{1...N}$ (angular coordinates and anatomical classes for the *N* nodes)
OUTPUT: $score_d$ (circular separation score based on distances)

the nodes are ranked according to the angular coordinates $\theta_{1...N}$ and ranks $r_{1...N}$ are assigned.
for each anatomical class $c$
    let $r^c_{1...N^c}$ be the subset of ranks for the $N^c$ nodes for which $C_{1...N} == c$ is satisfied.
    compute the pairwise circular distances $d(i,j)$ according to the following procedure:
    for $i = 1 ... N^c - 1$
        for $j = i + 1 ... N^c$
            $b = \max(r^c_i, r^c_j)$
            $a = \min(r^c_i, r^c_j)$
            $d(i,j) = \min(b - a, N - b + a)$
    compute the mean circular distance:
    $d^c_{mean} = \text{mean}(d(i,j))$
    using the same procedure, compute the mean circular distance $d^c_{worst}$ in the worst scenario in which the $N^c$ nodes are equidistantly arranged over the circle.
    using the same procedure, compute the mean circular distance $d^c_{best}$ in the best scenario in which the $N^c$ nodes are arranged in sequence over the circle without interruptions.
    compute the normalized $score^c_d$ for the anatomical class $c$:
$$score^c_d = \frac{d^c_{worst} - d^c_{mean}}{d^c_{worst} - d^c_{best}}$$
compute the final $score_d$ as the mean over the anatomical classes:
$score_d = \text{mean}_c (score^c_d)$

---

INPUT: $\theta_{1...N}$, $C_{1...N}$ (angular coordinates and anatomical classes for the *N* nodes)
OUTPUT: $score_w$ (circular separation score based on wrong nodes between the class extremes)

the nodes are ranked according to the angular coordinates $\theta_{1...N}$ and ranks $r_{1...N}$ are assigned.
for each anatomical class $c$
    let $r^c_{1...N^c}$ be the subset of ranks for the $N^c$ nodes for which $C_{1...N} == c$ is satisfied.
    sort the ranks $r^c_{1...N^c}$ obtaining $s^c_{1...N^c}$
    compute the number of nodes of a different class between consecutive nodes of the class $c$:
    for $i = 1 ... N^c - 1$
        $w(i) = s^c_{i+1} - s^c_i - 1$
    $w(N^c) = N - s^c_{N^c} + s^c_1 - 1$
    compute the total number of wrong nodes between the class extremes, which are the consecutive *c*-class nodes with the highest number of wrong nodes between them:
    $w^c = \underset{i=1...N^c}{\text{sum}}(w(i)) - \underset{i=1...N^c}{\max}(w(i))$
    compute the total number of wrong nodes between the class extremes $w^c_{worst}$ in the worst scenario in which the $N^c$ nodes are equidistantly arranged over the circle.
$$w^c_{worst} = \text{ceil}\left((N - N^c) * \frac{N^c - 1}{N^c}\right)$$
    in the best scenario in which the $N^c$ nodes are arranged in sequence over the circle without interruptions the total number of wrong nodes between the class extremes is 0.
compute the normalized final $score_w$ according to two different formulas:
$$score_{w1} = 1 - \underset{c}{\text{mean}}\left(\frac{w^c}{w^c_{worst}}\right)$$
$$score_{w2} = 1 - \frac{\underset{c}{\text{sum}}(w^c)}{\underset{c}{\text{sum}}(w^c_{worst})}$$

Note: since the two formulas gave results almost identical, we only report $score_{w1}$.

**Supplementary References**

1.  Gupta, A. *et al.* Sex Differences in the Influence of Body Mass Index on Anatomical Architecture of Brain Networks. *Int. J. Obes.* 1–46 (2017). doi:10.1038/ijo.2017.86
2.  Gupta, A. *et al.* Patterns of brain structural connectivity differentiate normal weight from overweight subjects. *NeuroImage Clin.* **7,** 506–517 (2015).
3.  Irimia, A., Chambers, M. C., Torgerson, C. M. & Van Horn, J. D. Circular representation of human cortical networks for subject and population-level connectomic visualization. *Neuroimage* **60,** 1340–1351 (2012).
4.  Labus, J. S. *et al.* Multivariate morphological brain signatures predict patients with chronic abdominal pain from healthy control subjects. *Pain* **8,** 1545–1554 (2015).
5.  Woodworth, D. *et al.* Unique Microstructural Changes in the Brain Associated with Urological Chronic Pelvic Pain Syndrome (UCPPS) Revealed by Diffusion Tensor MRI, Super-Resolution Track Density Imaging, and Statistical Parameter Mapping: A MAPP Network Neuroimaging Study. *PLoS One* **10,** e0140250 (2015).
6.  Destrieux, C., Fischl, B., Dale, A. & Halgren, E. Automatic parcellation of human cortical gyri and sulci using standard anatomical nomenclature. *Neuroimage* **53,** 1–15 (2010).
7.  Fischl, B. *et al.* Whole brain segmentation: Automated labeling of neuroanatomical structures in the human brain. *Neuron* **33,** 341–355 (2002).
8.  Mori, S., Crain, B. J., Chacko, V. P. & van Zijl, P. C. Three-dimensional tracking of axonal projections in the brain by magnetic resonance imaging. *Ann. Neurol.* **45,** 265–9 (1999).
9.  Hughes, A. J., Daniel, S. E., Kilford, L. & Lees, A. J. Accuracy of clinical diagnosis of idiopathic Parkinson's disease: a clinico-pathological study of 100 cases. *J. Neurol. Neurosurg. Psychiatry* **55,** 181–184 (1992).
10. Movement Disorder Society Task Force on Rating Scales for Parkinson's Disease. The Unified Parkinson's Disease Rating Scale (UPDRS): status and\rrecommendations. *Mov. Disord.* **18,** 738–750 (2003).
11. Hoehn, M. M. & Yahr, M. D. Parkinsonism: onset, progression, and mortality. *Neurology* **57,** 318 and 16 pages following (1967).
12. Nasreddine, Z. S. *et al.* The Montreal Cognitive Assessment, MoCA: A brief screening tool for mild cognitive impairment. *J. Am. Geriatr. Soc.* **53,** 695–699 (2005).
13. Beck, A. T., Ward, C. H., Mendelson, M., Mock, J. & Erbaugh, J. An inventory for measuring depression. *Arch. Gen. Psychiatry* **4,** 561–571 (1961).

14. Milardi, D. *et al.* Extensive Direct Subcortical Cerebellum-Basal Ganglia Connections in Human Brain as Revealed by Constrained Spherical Deconvolution Tractography. *Front. Neuroanat.* **10,** 29 (2016).
15. Cacciola, A. *et al.* A Direct Cortico-Nigral Pathway as Revealed by Constrained Spherical Deconvolution Tractography in Humans. *Front. Hum. Neurosci.* **10,** 374 (2016).
16. Cacciola, A. *et al.* Constrained Spherical Deconvolution Tractography Reveals Cerebello-Mammillary Connections in Humans. *Cerebellum* **16,** 483–495 (2017).
17. Milardi, D. *et al.* Red nucleus connectivity as revealed by constrained spherical deconvolution tractography. *Neurosci. Lett.* **626,** 68–73 (2016).
18. Milardi, D. *et al.* The Olfactory System Revealed: Non-Invasive Mapping by using Constrained Spherical Deconvolution Tractography in Healthy Humans. *Front. Neuroanat.* **11,** 1–11 (2017).
19. Besson, P. *et al.* Structural connectivity differences in left and right temporal lobe epilepsy. *Neuroimage* **100,** 135–144 (2014).
20. Desikan, R. S. *et al.* An automated labeling system for subdividing the human cerebral cortex on MRI scans into gyral based regions of interest. *Neuroimage* **31,** 968–980 (2006).
21. Fischl, B. *et al.* Sequence-independent segmentation of magnetic resonance images. in *NeuroImage* **23,** S69-84 (2004).
22. Smith, R. E., Tournier, J. D., Calamante, F. & Connelly, A. The effects of SIFT on the reproducibility and biological accuracy of the structural connectome. *Neuroimage* **104,** 253–265 (2015).
23. Patenaude, B., Smith, S. M., Kennedy, D. N. & Jenkinson, M. A Bayesian model of shape and appearance for subcortical brain segmentation. *Neuroimage* **56,** 907–922 (2011).
24. Gong, G. *et al.* Mapping anatomical connectivity patterns of human cerebral cortex using in vivo diffusion tensor imaging tractography. *Cereb. Cortex* **19,** 524–536 (2009).
25. Zhang, R. *et al.* Disrupted brain anatomical connectivity in medication-naïve patients with first-episode schizophrenia. *Brain Struct. Funct.* **220,** 1145–1159 (2015).
26. Hagmann, P. *et al.* Mapping human whole-brain structural networks with diffusion MRI. *PLoS One* **2,** e597 (2007).
27. Iturria-Medina, Y., Sotero, R. C., Canales-Rodríguez, E. J., Alemán-Gómez, Y. & Melie-García, L. Studying the human brain anatomical network via diffusion-weighted

MRI and Graph Theory. *Neuroimage* **40,** 1064–1076 (2008).

28. Zalesky, A. *et al.* Whole-brain anatomical networks: Does the choice of nodes matter? *Neuroimage* **50,** 970–983 (2010).